\documentclass[a4paper]{article}

\usepackage{INTERSPEECH2021}
\usepackage{amsmath,graphicx,array,multirow}
\title{Inplace Gated Convolutional Recurrent Neural Network \\ For Dual-channel Speech Enhancement
 }
\name{Jinjiang Liu, Xueliang Zhang}

\address{
College of Computer Science, Inner Mongolia University, China}
\email{jetliu1994@foxmail.com,cszxl@imu.edu.cn}

\begin{document}

\maketitle
\begin{abstract}
For dual-channel speech enhancement, it is a promising idea to
design an end-to-end model based on the traditional array signal
processing guideline and the manifold space of multi-channel
signals. We found that the idea above can be effectively implemented by the classical convolutional recurrent neural networks (CRN) architecture. We propose a very compact inplace
gated convolutional recurrent neural network (inplace GCRN)
for end-to-end multi-channel speech enhancement, which utilizes inplace-convolution for frequency pattern extraction and
reconstruction. The inplace characteristics efficiently preserve
spatial cues in each frequency bin for channel-wise long short-term memory neural networks (LSTM) tracing the spatial source.
In addition, we come up with a new spectrum recovery method
by predict amplitude mask, mapping, and phase, which effectively improves the speech quality.

\end{abstract}


\section{Introduction}
\label{sec:intro}

Speech enhancement is very important to many real applications, such as telecommunication, robust automatic speech recognition (ASR), and hearing aids.
For better speech quality and intelligibility, most of the devices, e.g. mobile phone and smart home device, are equipped with multiple microphones which can utilize spatial information.
Dual-microphone is the most common configuration.

Traditionally, multi-channel speech enhancement can be divided into two categories.
One is the blind source separation (BSS) method \cite{bbs1}\cite{bbs2},
which is under the assumption of the independence of source signals.
BSS-based speech enhancement separates signals by adaptively optimizing the cost function of the independent components analysis (ICA) process.

The other one is beamforming \cite{BF1}\cite{BF2}\cite{GSC_dc}, which utilizes the direction of arrival (DOA) and second-order statistics of signals.

Recently, deep learning has achieved great progress in multi-channel speech enhancement.
Generally, the deep-learning-based methods can be divided into two categories.
One way is to combine deep learning with the traditional methods.
The representative method is mask-based beamforming \cite{mask_BF} \cite{mask_BF2}, which calculated beamformer coefficients with the help of a mask estimated by deep neural networks (DNN).

Instead of estimating mask, Wang and Wang used a deep neural network to directly estimate the complex spectral which is utilized to computing a minimum variance distortion-less response (MVDR) beamformer \cite{cs_MVDR}.
Zhang and Wang used spectral features extracted by fixed beamforming and spatial features as the input of a DNN for binaural speech enhancement \cite{xueliangzhang}.
Li et al. used two fixed differential beamformers with opposite directions as a robust discriminative feature for the neural network to directly estimate the amplitude mask \cite{BF_feat}.

Another is the full neural network-based or end-to-end method.
Wang and Wang proposed an all-neural multi-channel speech enhancement \cite{all_NN}.
Tan et al. utilized a convolutional recurrent network for dual-microphone speech enhancement \cite{crn_mobile}.
Gu et al. proposed an end-to-end network architecture for multi-channel speech separation in the time domain, which aims to learn spatial information directly from multi-channel waveform instead of widely-used Short Time-Frequency Transform (STFT) \cite{BF_feat2}.
Most of these algorithms mentioned above are finely designed with multistage training or process. However, separately trained modules may not cooperate well, because the hand-crafted interface may lead to information distortions and limits the ability of neural networks. Instead, a well-designed end-to-end system naturally fits the solution's manifold space of the original task.

\begin{figure}[t]
  \centering
  \includegraphics[width=\linewidth]{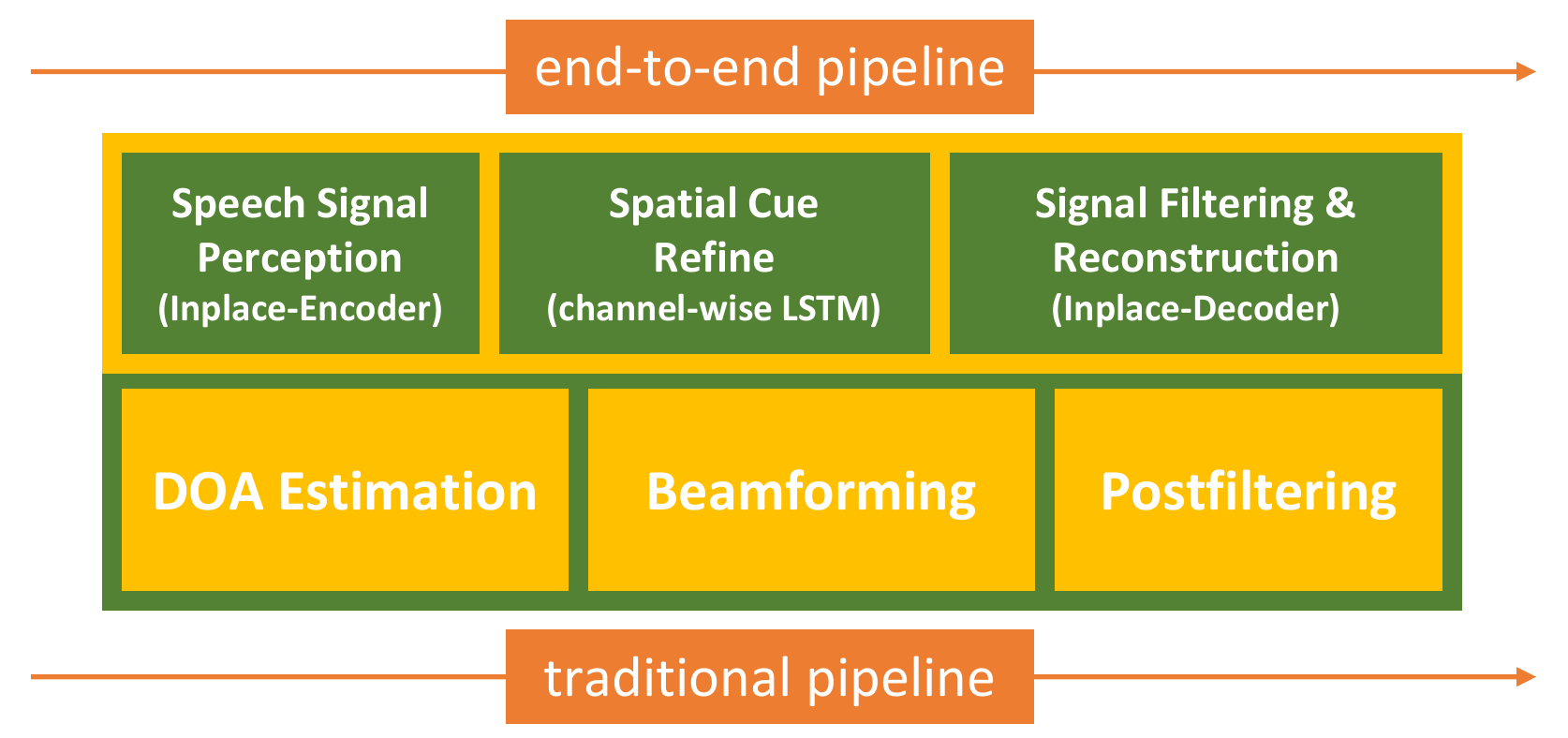}
  \caption{inplace GCRN based end-to-end speech enhancement pipeline comparing module functioning with traditional method pipeline.}
\vspace{-5mm}
\end{figure}

Inspired by the three steps of beamforming technique, DOA estimation, beamforming, and post-filtering, we propose end-to-end dual-channel speech enhancement, as shown in Figure 1.
The pipeline consists of speech signal perception, spatial cue processing, and speech signal reconstruction, which are implemented by the similar architecture of CRN \cite{gcrn_mapping}.
It should be mentioned that these three steps don't exactly correspond to the traditional array speech processing pipeline due to its end-to-end nature.
The typical CRN utilizes the convolution with stride operation to shrink and expand the feature on the frequency dimension in the encoder and decoder stage, respectively.
However,  wideband beamforming is processed in each frequency bin independently.
And we call this inplace process. Therefore, we propose an inplace GCRN model which is consists of an inplace-encoder, channel-wise LSTM shared by all frequency bin, and inplace-decoder. Experimental results show that the proposed inplace GCRN can dramatically improve the performance.

The paper is organized as follows.
In Section 2, we describe the core ideas and show key details of the inplace GCRN model and feature design.
In Section 3, we show the setup and details of the experiment, the experimental result, and the analysis.
We make conclusions in Section 4.

\vspace{2mm}
\section{Algorithm}

For a dual-channel microphone array system, the received signal $x_m(k)$ can be modeled as follows:

\begin{equation}
\footnotesize
\begin{split}
x_{m}(k) = s(k) *h_{s,m}(k) + n(k)*h_{n,m}(k) \\
\end{split}
\vspace{1mm}
\end{equation}
where, $m$ denotes the channel number, $ s(k) $ and $ n(k) $ indicate speech signal and noise signal.
$ h_{s,m}(k) $ and $ h_{n,m}(k) $ are the acoustic impulse response from speech source and noise source to $m$-th microphone, respectively,
and '$*$' is the convolution operation.

\subsection{Inplace GCRN}

The inplace GCRN is mainly constructed by inplace convolution gated linear unit (GLU) and channel-wise LSTM to analyzing noisy input features and synthesize clean speech features.

\begin{figure}[h]
  \centering
  \includegraphics[width=\linewidth]{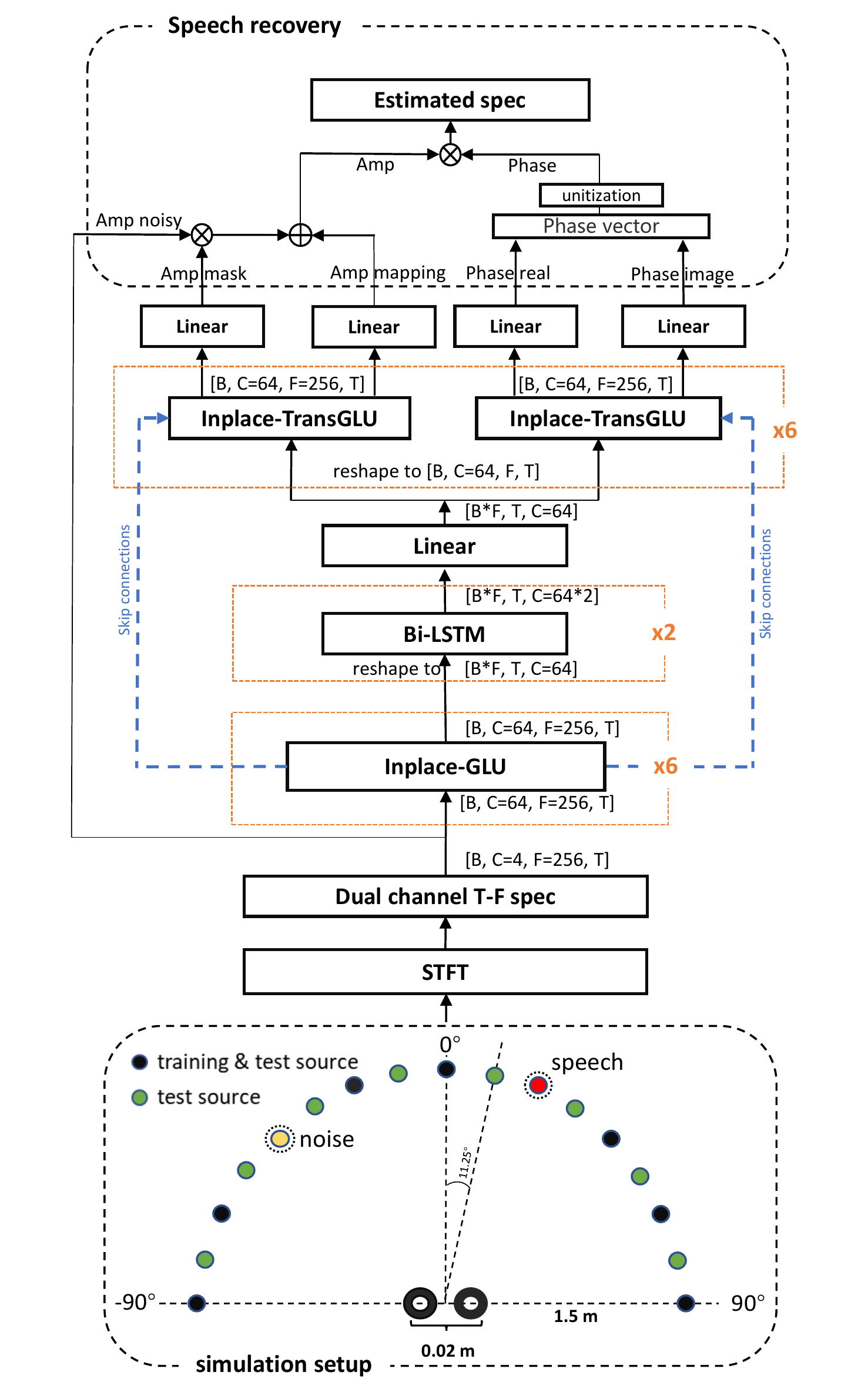}
  \caption{The proposed dual-channel speech enhancement system and the room simulation setup.}
\end{figure}

\subsubsection{Inplace convolution}

Inplace-convolution is the convolutional neural network that the stride of the convolving kernel is set to one.
It means that the inplace convolution does not downsample the features in the frequency dimension. In this way, the spatial correlations are naturally and explicitly maintained in each frequency bin.
In the conventional CRN structure, the stride of convolutional operation in the frequency dimension is normally set to 2, which shrinks the feature in the frequency dimension.
By stacking the convolutional layers several times, the patterns lying in the frequency dimension are encoded into the channel dimension.
This is very effective for the single-channel task to model speech harmonic structural patterns and tracking their variations in the time domain.
But for multi-channel speech enhancement,
the downsampling convolution aliases spatial cues with speech patterns in channel dimension, which makes later LSTM hard to extract the spatial information.

\subsubsection{Channel-wise LSTM with model reuse mechanism}
The conventional CRN model using the LSTM model to process overall frequency bins.
In contrast, we apply LSTMs on each frequency bin, the input feature is only containing channel-wise features without frequency dimension.

Due to the inplace characteristic of the encoder, the spatial cue will be explicitly maintained inside each frequency bin, without being obscured with its neighborhoods by the encoding process on the frequency dimension.
So the processing of extract spatial information for each frequency bin could be done independently, which is similar to the beamforming method.
There is one thing different comparing to the beamforming method, due to the difference of wavelength in the different frequency bands, if we want to make a same phase compensation to form a same beam pattern for different frequencies, the beamformer weight for each band is different.
But the LSTM model does not pick up speech by phase compensation, it only needs to analyze spatial information by time delay, the time delay for a certain look direction in different frequency bins is the same, so we could process all the frequency bins by reusing one LSTM model. This LSTM reuse mechanism makes the whole model very compact.

\subsection{Amplitude and phase prediction}
For phase prediction, Yin et al. \cite{phasen} show that it has benefits to estimate the amplitude and phase separately, compared with complex ratio mask \cite{CRM}.
When it comes to amplitude prediction, mask and mapping are two common ways. Estimating mask works well for high SNR conditions due to it can directly use the input features, while the mapping method performs better, in low SNR conditions.
The characteristics of the spectrogram recovered by them could be different and somewhere complementary.
Zhang et al.\cite{mask_mapping} use two networks to predict the amplitude mask and amplitude itself respectively, and then, use another network to combine the outputs of the two models to achieve better performance.
In our model, two decoders are used to separately predict the amplitude and phase, masking and mapping are done by a single decoder, and another decoder is used to predict the phase.

\subsection{System construction}
The proposed system is shown in Figure 2.
We use the short-time Fourier transform (STFT) to extract the complex spectrum of two channels, and concatenate their real and imaginary parts as channel dimensions of the input features of the model.
The input feature in the shape of [batch, channel=4, frequency, time] is first processed by six cascades 5x1 kernel inplace GLU, which is constructed by inplace convolution as follow:
\begin{equation}
\footnotesize
Y = ELU(BN(iConv(X)\otimes Sigmoid(iConv(X))))
\end{equation}
where $ELU(.)$ and $Sigmoid(.)$ are the activation functions, $BN(.)$ is the batch normalization, the $iConv$  is the inplace convolution, $ \otimes$ denotes for the element-wise multiplication.

After the encoder, we use channel-wise LSTM to refine the spatial information.
That is, technically, we merge the frequency dimension of the encoder's output feature to the batch dimension through reshape operation as [batch x frequency, time,channel=64], and put it into a Bi-LSTM model with two layers and 64 feature size.
After the Bi-LSTM, the output feature is passing a linear layer to half its channel number and reshapes back, finally the feature is duplicated as two decoders' input.

The decoder is constructed by six inplace cascades Transpose GLU, the Transpose GLU is defined as follow:
\begin{equation}
\footnotesize
Y = ELU(BN(iTConv(X)\otimes Sigmoid(iTConv(X))))
\end{equation}
where $iTConv$ is the inplace transpose convolution.
The $i$ th GLU's output concatenated with $i-1$ th transpose GLU's output as the $i$ th transpose GLU's input, this lead to the skip connections.
The input channel of transpose GLU is 128, the output channel for both GLU and transpose GLU is constantly 64, except the output layer of decoders.
A more detailed description of our proposed network hyperparameters is provided in Table 1.

There are two channels of output from both amplitude decoder and phase decoder, every output features passing a linear layer with 256 unit as final output feature, two output of amplitude decoder predict amplitude mask and amplitude mapping.
We generate the estimated amplitude spectrogram and phase spectrogram as follow:
\vspace{0mm}
\begin{footnotesize}
\begin{align}
\footnotesize
& A_{est} = A_{msk} \otimes A_{nsy}+A_{map}  \\
& P_{est} = \frac{P_{est_{r}}+jP_{est_{i}}}{\sqrt{P_{est_{r}}^2+P_{est_{i}}^2}}  \\
& X_{est} = A_{est} \otimes P_{est}
\end{align}
\end{footnotesize}
where, $A_{msk}$ and $A_{map}$ are two outputs of amplitude decoder used for amplitude mask and amplitude mapping, $A_{nsy}$ is the noisy speech amplitude. $P_{est_{r}}$ and $P_{est_{i}}$ are two outputs of phase decoder used as the real and imaginary of phase $P_{est}$.
$ X_{est} $ denotes for the estimated complex spectrogram.

In the training stage, we use the Phasen loss function \cite{phasen}:
\vspace{-3mm}

\begin{scriptsize}
\begin{equation}
\begin{aligned}
&&L = & \frac{1}{F} \sum_{i=1}^{F} ((A_{s}[i])^{\frac{1}{3}}-(A_{est}[i])^{\frac{1}{3}})^2 + \\
&&    & \frac{1}{F} \sum_{i=1}^{F} ((A_{s}[i])^{\frac{1}{3}} \otimes P_{s_{r}}[i] - (A_{est}[i])^{\frac{1}{3}}\otimes P_{est_{r}}[i])^2 + \\
&&    & \frac{1}{F} \sum_{i=1}^{F} ((A_{s}[i])^{\frac{1}{3}} \otimes P_{s_{i}}[i] - (A_{est}[i])^{\frac{1}{3}}\otimes P_{est_{i}}[i])^2
\end{aligned}
\end{equation}
\end{scriptsize}
where $i$ is the index of frequency bin elements, and $F$ is the total number of frequency bins. $A_s$, $P_{s_{r}}$ and $P_{s_{i}}$ are amplitude, real and imaginary part of phase of clean speech spectrogram, respectively.

\begin{table}[htbp]
\scriptsize
  \centering
  \caption{ Architecture of our proposed IGCRN. T denotes the
number of time frames, B is the batch size.}
    \begin{tabular}{|c|c|c|c|}
    \hline
    layer name & input size & hyperparameters & output size \\
    \hline
    iGLU1 & [B,2,256,T] & 5x1, (1,1), 64 & [B,64,256,T] \\
    \hline
    iGLU2 $\sim$ 6 & [B,64,256,T] & 5x1, (1,1), 64 & [B,64,256,T] \\
    \hline
    reshape & [B,64,256,T] &       & [Bx256,T,64] \\
    \hline
    B-LSTM(2layer) & [Bx256,T,64] & 64    & [Bx256,T,128] \\
    \hline
    linear & [Bx256,T,128] & (128,64) & [Bx256,T,64] \\
    \hline
    reshape & [Bx256,T,64] &       & [B,64,256,T] \\
    \hline
    iTGLU6 $\sim$ 2 & [B,128,256,T] & 5x1, (1,1), 64 & [B,64,256,T] \\
    \hline
    iTGLU1 & [B,128,256,T] & 5x1, (1,1), 64 & [B,2,256,T] \\
    \hline
    \end{tabular}%
  \label{tab:addlabel}%
\vspace{0mm}
\end{table}%

\section{Experiment and Evaluation}
\subsection{Experimental setup}
For the speech corpus, we randomly select 29 hours and 1 hour of speech from the mandarin dataset AISHELL-1 \cite{aishell-1} as training and validation sets, respectively. To evaluate the generalization ability, we selected a 1-hour speech corpus from the TIMIT dataset for the test. The noises are from NOISEX92. We choose destroyerops, white, and babble for the test, and the remaining 12 noises are used for training.
We simulate room impulse response (RIR) by the IMAGE method\cite{rir_gen}.
Specifically, two microphones with $2cm$ interval are placed at the center of a $5m (length) \times 5m (width) \times 3m (height)$ room.
We use 9 source positions for training which are placed at $1.5m$ away from the center of the two microphones and ranged from $-90^{\circ}$ to $90^{\circ}$ with $22.5^{\circ}$ interval. Another 17 different positions are used for testing which are placed at the same distance with $11.25^{\circ}$ interval.
For each mixture, we first randomly choose a speech and a slice of noise, and then place them at two different positions, and mix the speech and noise at selected SNR, -3dB, 0dB, and 3dB.
The frame length is 32 ms and the frameshift 16 ms. The Square-root Hann window is used as the analysis window. The sampling rate is 16 kHz. A 512-point discrete Fourier transform is used to
extract complex STFT spectrograms.

All models are trained using Adam optimizer with a fixed learning rate of 0.0002, the minibatch is setting to 4. The detailed structure of the proposed inplace GCRN is shown in Figure 2.

\subsection{Experimental result}

In this study, short-time objective intelligibility (STOI) \cite{stoi}, perceptual evaluation of speech quality (PESQ) \cite{pesq}, and signal-to-distortion ratios (SDR) are employed as the evaluation metrics.
The best results in each case are highlighted by boldface.

\begin{table*}[!tbp]
  \caption{Comparisons of different approaches in terms of STOI, PESQ, and SDR in -3dB, 0dB, and 3dB direction noise.}
  \centering

  \footnotesize
\begin{tabular}{cc|ccc|ccc|ccc}
\hline
                &              & \multicolumn{3}{c|}{STOI}   & \multicolumn{3}{c|}{PESQ}   & \multicolumn{3}{c}{SDR}     \\ \hline
         SNR          & method & white & destroyerops & babble & white & destroyerops & babble & white & destroyerops & babble \\ \hline
\multirow{4}{*}{3dB}  & noisy & 0.78  & 0.73       & 0.71   & 1.71  & 1.93       & 1.9    & 3   & 3        & 3      \\
                      & MVDR  & 0.88  & 0.87       & 0.87   & 2.63  & 2.59       & 2.65   & 11.6  & 11.1       & 11.7   \\
                      & GCRN  & 0.90  & 0.90       & 0.91   & 2.71  & 2.81       & 2.91   & 9.3   & 9.1        & 9.0      \\
                      & IGCRN & \textbf{0.97}  & \textbf{0.98}       & \textbf{0.98}   & \textbf{3.75}  & \textbf{3.96}       & \textbf{3.95}   & \textbf{19.6}  & \textbf{21.6}       & \textbf{21.4}   \\ \hline
\multirow{4}{*}{0dB}  & noisy & 0.71  & 0.67       & 0.65   & 1.49  & 1.7        & 1.69   & 0     & 0          & 0      \\
                      & MVDR  & 0.87  & 0.85       & 0.85   & 2.55  & 2.51       & 2.54   & 8.6   & 7.8        & 8.4    \\
                      & GCRN  & 0.88  & 0.89       & 0.89   & 2.57  & 2.75       & 2.79   & 6.3   & 6.5        & 6.1    \\
                      & IGCRN & \textbf{0.96}  & \textbf{0.97}       & \textbf{0.97}   & \textbf{3.59}  & \textbf{3.87}       & \textbf{3.89}   & \textbf{18.4}  & \textbf{20.6}       & \textbf{20.5}   \\ \hline
\multirow{4}{*}{-3dB} & noisy & 0.64  & 0.61       & 0.58   & 1.29  & 1.46       & 1.49   & -3    & -3         & -3     \\
                      & MVDR  & 0.85  & 0.84       & 0.83   & 2.49  & 2.45       & 2.46   & 5.4   & 4.4        & 5.3    \\
                      & GCRN  & 0.85  & 0.84       & 0.85   & 2.35  & 2.54       & 2.59   & 3.4   & 3.5        & 3.3    \\
                      & IGCRN & \textbf{0.94}  & \textbf{0.95}       & \textbf{0.96}   & \textbf{3.36}  & \textbf{3.68}       & \textbf{3.75}   & \textbf{15.6}  & \textbf{18.6}       & \textbf{19.2}   \\ \hline

\end{tabular}
\vspace{0mm}
\end{table*}

First, we compare the proposed IGCRN with the conventional beamformer, MVDR, and the gated CRN in different noisy conditions at different SNR. It should be mentioned that the true direction is given for MVDR, which has to be estimated in practice. The results are shown in Table 2. It can be seen that the proposed IGCRN  significantly and consistently outperforms the comparison methods in all conditions. The average STOI and PESQ gains are over $30\%$ and $2.0$ compare to the unprocessed noisy speech.

\begin{table*}[!tbp]
  \caption{Comparisons of different approaches in terms of STOI, PESQ, and SDR in -3dB direction noise.}
  \centering
  \footnotesize
\begin{tabular}{c|ccc|ccc|ccc}
\hline
                & \multicolumn{3}{c|}{STOI}   & \multicolumn{3}{c|}{PESQ}   & \multicolumn{3}{c}{SDR}     \\ \hline
     method     & white & destroyerops & babble & white & destroyerops & babble & white & destroyerops & babble \\ \hline
noisy            & 0.64  & 0.61       & 0.58   & 1.29  & 1.46       & 1.49   & -3    & -3         & -3     \\
GCRN(CS)         & 0.85  & 0.84       & 0.85   & 2.35  & 2.54       & 2.59   & 3.4   & 3.5        & 3.3    \\
GCRN(Msk+Ps)     & 0.90  & 0.87       & 0.85   & 2.74  & 2.75       & 2.62   & 11.6  & 10         & 8.5    \\
GCRN(Msk+Map+Ps) & 0.90  & 0.88       & 0.87   & 2.89  & 2.87       & 2.77   & 11.8  & 11.3       & 10.4   \\
IGCRN            & \textbf{0.94}  & \textbf{0.95}       & \textbf{0.96}   & \textbf{3.36}  & \textbf{3.68}       & \textbf{3.75}   & \textbf{15.6}  & \textbf{18.6}       & \textbf{19.2}   \\ \hline
\end{tabular}
\vspace{0mm}
\end{table*}

Another contribution of this work is the proposed training target.
In order to evaluate the effectiveness, we compare the performances of GCRN with different outputs. The results are shown in Table 3, where GCRN(CS) is the original complex spectral mapping, GCRN(Msk+Ps) is to estimate the amplitude mask and clean phase, and GCRN(Msk+Mp+Ps) is the proposed target. The results are shown in Table 3.
It can be seen that compared with the original GCRN, the effect of predicting mask and phase is better.
It is because amplitude is more important than phase, and the amplitude and phase are coupled in the complex spectrum. Similar results are observed in \cite{wangzhongqiu1}\cite{wangzhongqiu2}, where both complex and magnitude spectrum are restrained.
For GCRN(Msk+Map+Ps) we the introduced amplitude mapping term can further improve the performance, which pays more attention to the amplitude of spectrum than the others. However, GCRN(Msk+Map+Ps) is still much worse than the proposed IGCRN.

\setlength{\tabcolsep}{0.75mm}{
\begin{table*}[!t]
  \caption{Comparisons of different methods in terms of different DOA with $11^{\circ}$ degree included angle of speech and noise in -3 dB babble direction noise, S and N are the DOA of speech and noise respectively.}
  \centering
  \footnotesize
\begin{tabular}{c|ccc|ccc|ccc}
  \hline
       & \multicolumn{3}{c|}{STOI(0.58)} & \multicolumn{3}{c|}{PESQ(1.49)} & \multicolumn{3}{c}{SDR(-3)} \\ \hline
   DOA & MVDR     & GCRN(Msk+Map+Ps)     & IGCRN     & MVDR     & GCRN(Msk+Map+Ps)     & IGCRN     & MVDR    & GCRN(Msk+Map+Ps)    & IGCRN    \\ \hline
$S=0^{\circ}, \ \  N=11^{\circ}$ & 0.87     & 0.85     & \textbf{0.95}      & 2.74     & 2.64     & \textbf{3.54}      & 11.6    & 9.2     & \textbf{17.5}     \\
$S=23^{\circ},N=34^{\circ}$ & 0.76     & 0.74     & \textbf{0.94}      & 2.37     & 2.16     & \textbf{3.40}      & 6.4     & 3.7     & \textbf{15.4}     \\
$S=45^{\circ},N=56^{\circ}$ & 0.69     & 0.66     & \textbf{0.89}      & 1.80     & 1.89     & \textbf{2.86}      & 0.2     & 1.3     & \textbf{9.7 }     \\
$S=68^{\circ},N=79^{\circ}$ & 0.60     & 0.61     & \textbf{0.73}      & 1.55     & 1.74     & \textbf{2.17}      & -1.7    & 0.1     & \textbf{3.6 }     \\
$S=79^{\circ},N=90^{\circ}$ & 0.54     & \textbf{0.58}     &        {0.57}      & 1.34     & 1.69     & \textbf{1.70}      & -6.4    & -0.5    & \textbf{0.4 }     \\ \hline
\end{tabular}
\vspace{-1mm}
\end{table*}
}

It is known that the spatial information is reflected by the time delay between the two microphones. The resolution of time delay is non-uniform to the directions. So, we investigate the performance of the methods when the target speech comes from different directions. In Table 4, it can be seen that performances gradually decay when the direction moves from $0^\circ$ to $90^\circ$, because the difference between the time delays of speech and noise becomes small. Compared with MVDR, GCRN is not good in high-resolution conditions, e. g. $S=0^\circ$ and $23^\circ$. However, in low-resolution conditions, GCRN outperforms the MVDR, because GCRN utilizes both spectral and spatial information. However, the proposed IGCRN outperforms the MVDR and GCRN in all the conditions. It implies that IGCRN can make better use of spatial information than GCRN.

\setlength{\tabcolsep}{2mm}{
\begin{table}[!t]

  \caption{Investigation of influence of the downsampling in -3dB babble noise condition.}
  \centering
  \scriptsize
\linespread{1}

    \begin{tabular}{lccccc}
\hline
    method   & STOI  & PESQ  & MAC(G) & Params(M) & LSTM \\
\hline
    noisy & 0.583  & 1.49  &       &       &  \\

    GCRN            & 0.847 & 2.59  & 28.8  & 71.8  & 1024 \\
    IGCRN64         & 0.968 & 3.83  & 19.9  & 1.4   & 64 \\
    IGCRN80         & \textbf{0.982} & \textbf{4.02}  & 31.1  & 2.3   & 80 \\
    IGCRN64-1DS     & 0.982 & 3.94  & 32.1  & 3.5   & 128 \\
    IGCRN64-2DS     & 0.981 & 3.91  & 53.3  & 9.5   & 256 \\
    IGCRN64-3DS     & 0.974 & 3.73  & 85.3  & 24.1  & 512 \\
    IGCRN64-4DS     & 0.961 & 3.58  & 149.5 & 82.5  & 1024 \\
    IGCRN64-5DS     & 0.954 & 3.52  & 277.8 & 316.3 & 2048 \\
    IGCRN64-6DS     & 0.949 & 3.51  & 430.8 & 777.3 & 2048 \\
\hline
    \end{tabular}%
  \label{tab:addlabel}%
\vspace{-2mm}
\end{table}%
}

In Table 5, we show that how the downsampling operation affects the performance, where multiply-accumulate operations (MAC) and total trainable parameters (Params) are also listed.
For IGCRN(n)-(k)DS, n and k denote the number of the channel of the first GLU output feature and the times of downsampling in convolution layers.
For each downsampling operation, we will double the channel dimension of its output feature.
We expand the IGCRN channel from the original 64 to 80, so the MAC of IGCRN80 is similar to the 1DS model.
From Table 5, we can see that the performance gradually drops when the downsampling operation increasing, even though the complexity of the model is significantly increased.
This result shows the importance of the inplace characteristic when we doing multi-channel enhancement in the time-frequency domain.

When it comes to parameter efficiency, the reuse mechanism of channel-wise LSTM makes the inplace GCRN model extremely compact with only 1.4 million parameters, and also the computational complexity is lower than the conventional GCRN model.

\section{Conclusions}
In this study, we propose a compact inplace GCRN model for dual-channel enhancement. Experimental results show that the proposed method can effectively exploit and utilize the spatial source information, which is guaranteed by the inplace characteristics of the inplace GCRN model, and it reveals the huge potential of designing a proper neural network for a certain task with a specific sparse manifold space.

\vspace{1mm}
\section{Acknowledgements}
This research is supported by the National Natural Science Foundation of China (No. 61876214).

\newpage

\bibliographystyle{IEEEtran}

\begin{thebibliography}{10}
\providecommand{\url}[1]{#1}
\csname url@samestyle\endcsname
\providecommand{\newblock}{\relax}
\providecommand{\bibinfo}[2]{#2}
\providecommand{\BIBentrySTDinterwordspacing}{\spaceskip=0pt\relax}
\providecommand{\BIBentryALTinterwordstretchfactor}{4}
\providecommand{\BIBentryALTinterwordspacing}{\spaceskip=\fontdimen2\font plus
\BIBentryALTinterwordstretchfactor\fontdimen3\font minus
  \fontdimen4\font\relax}
\providecommand{\BIBforeignlanguage}[2]{{%
\expandafter\ifx\csname l@#1\endcsname\relax
\typeout{** WARNING: IEEEtran.bst: No hyphenation pattern has been}%
\typeout{** loaded for the language `#1'. Using the pattern for}%
\typeout{** the default language instead.}%
\else
\language=\csname l@#1\endcsname
\fi
#2}}
\providecommand{\BIBdecl}{\relax}
\BIBdecl

\bibitem{bbs1}
D.~Mohamed and R.~Bendoumia, ``A new adaptive filtering subband algorithm for
  two-channel acoustic noise reduction and speech enhancement,''
  \emph{Computers and Electrical Engineering}, vol.~39, p. 2531–2550, 2013.

\bibitem{bbs2}
D.~Mohamed, A.~Gilloire, and P.~Scalart, ``Noise cancellation using two closely
  spaced microphones: Experimental study with a specific model and two adaptive
  algorithms,'' in \emph{IEEE International Conference on Acoustics, Speech and
  Signal Processing (ICASSP)}, vol.~3, 2006.

\bibitem{BF1}
Applebaum, S., Chapman, and D., ``Adaptive arrays with main beam constraints,''
  \emph{IEEE Transactions on Antennas and Propagation}, vol.~24, no.~5, pp.
  650--662, 1976.

\bibitem{BF2}
J.~Capon, ``High-resolution frequency-wavenumber spectrum analysis,''
  \emph{Proceedings of the IEEE}, vol.~57, no.~8, pp. 1408--1418, 1969.

\bibitem{GSC_dc}
K.~M. Buckley and L.~J. Griffiths, ``An adaptive generalized sidelobe canceller
  with derivative constraints,'' \emph{IEEE Transactions on Antennas and
  Propagation}, vol.~34, no.~3, pp. 311--319, 1986.

\bibitem{mask_BF}
J.~{Heymann}, L.~{Drude}, and R.~{Haeb-Umbach}, ``Neural network based spectral
  mask estimation for acoustic beamforming,'' in \emph{IEEE International
  Conference on Acoustics, Speech and Signal Processing (ICASSP)}, 2016, pp.
  196--200.

\bibitem{mask_BF2}
J.~Heymann, M.~Bacchiani, and T.~Sainath, ``Performance of mask based
  statistical beamforming in a smart home scenario,'' in \emph{IEEE
  International Conference on Acoustics, Speech and Signal Processing
  (ICASSP)}, 2018, pp. 6722--6726.

\bibitem{cs_MVDR}
Z.-Q. Wang, P.~Wang, and D.~Wang, ``Complex spectral mapping for single-and
  multi-channel speech enhancement and robust asr,'' \emph{IEEE/ACM
  Transactions on Audio, Speech, and Language Processing}, vol.~28, pp.
  1778--1787, 2020.

\bibitem{xueliangzhang}
X.~Zhang and D.~L. Wang, ``Deep learning based binaural speech separation in
  reverberant environments,'' \emph{IEEE/ACM Transactions on Audio Speech and
  Language Processing}, vol.~25, no.~5, pp. 1075--1084, 2017.

\bibitem{BF_feat}
H.~Li, X.~Zhang, and G.~Gao, ``Beamformed feature for learning-based
  dual-channel speech separation,'' in \emph{IEEE International Conference on
  Acoustics, Speech and Signal Processing (ICASSP)}, 2020, pp. 4722--4726.

\bibitem{all_NN}
\BIBentryALTinterwordspacing
Z.-Q. Wang and D.~Wang, ``All-neural multi-channel speech enhancement,'' in
  \emph{Interspeech}, 2018, pp. 3234--3238. [Online]. Available:
  \url{http://dx.doi.org/10.21437/Interspeech.2018-1664}
\BIBentrySTDinterwordspacing

\bibitem{crn_mobile}
K.~Tan, X.~Zhang, and D.~Wang, ``Real-time speech enhancement using an
  efficient convolutional recurrent network for dual-microphone mobile phones
  in close-talk scenarios,'' in \emph{IEEE International Conference on
  Acoustics, Speech and Signal Processing (ICASSP)}.\hskip 1em plus 0.5em minus
  0.4em\relax IEEE, 2019, pp. 5751--5755.

\bibitem{BF_feat2}
R.~{Gu}, S.~{Zhang}, L.~{Chen}, Y.~{Xu}, M.~{Yu}, D.~{Su}, Y.~{Zou}, and
  D.~{Yu}, ``Enhancing end-to-end multi-channel speech separation via spatial
  feature learning,'' in \emph{IEEE International Conference on Acoustics,
  Speech and Signal Processing (ICASSP)}, 2020, pp. 7319--7323.

\bibitem{gcrn_mapping}
K.~Tan and D.~L. Wang, ``Learning complex spectral mapping with gated
  convolutional recurrent networks for monaural speech enhancement,''
  \emph{IEEE/ACM Transactions on Audio, Speech, and Language Processing},
  vol.~28, pp. 380--390, 2019.

\bibitem{phasen}
D.~Yin, C.~Luo, Z.~Xiong, and W.~Zeng, ``Phasen: A phase-and-harmonics-aware
  speech enhancement network,'' \emph{Proceedings of the AAAI Conference on
  Artificial Intelligence}, vol.~34, pp. 9458--9465, 2020.

\bibitem{CRM}
D.~S. {Williamson} and D.~{Wang}, ``Time-frequency masking in the complex
  domain for speech dereverberation and denoising,'' \emph{IEEE/ACM
  Transactions on Audio, Speech, and Language Processing}, vol.~25, no.~7, pp.
  1492--1501, 2017.

\bibitem{mask_mapping}
H.~Zhang, X.~Zhang, and G.~Gao, ``Multi-target ensemble learning for monaural
  speech separation,'' in \emph{Interspeech}, 2017.

\bibitem{aishell-1}
H.~Bu, J.~Du, X.~Na, B.~Wu, and H.~Zheng, ``Aishell-1: An open-source mandarin
  speech corpus and a speech recognition baseline,'' in \emph{Conference of The
  Oriental Chater of International Committee for Coordination and
  Standardization of peech Databases and Assessment Techniques}.

\bibitem{rir_gen}
E.~Habets, ``Room impulse response generator,'' pp. 1--17, 2006.

\bibitem{stoi}
C.~H. Taal, R.~C. Hendriks, R.~Heusdens, and J.~Jensen, ``An algorithm for
  intelligibility prediction of time–frequency weighted noisy speech,''
  \emph{IEEE Transactions on Audio and Speech Language Processing}, vol.~19,
  no.~7, pp. 2125--2136, 2011.

\bibitem{pesq}
A.~W. Rix, J.~G. Beerends, M.~P. Hollier, and A.~P. Hekstra, ``Perceptual
  evaluation of speech quality (pesq)-a new method for speech quality
  assessment of telephone networks and codecs,'' in \emph{IEEE International
  Conference on Acoustics}, 2002.

\bibitem{wangzhongqiu1}
Z.~Q. {Wang} and D.~{Wang}, ``Deep learning based target cancellation for
  speech dereverberation,'' \emph{IEEE/ACM Transactions on Audio, Speech, and
  Language Processing}, vol.~28, pp. 941--950, 2020.

\bibitem{wangzhongqiu2}
Z.~Q. {Wang} and D.~{Wang}, ``Multi-microphone complex spectral mapping for speech
  dereverberation,'' in \emph{IEEE International Conference on Acoustics,
  Speech and Signal Processing (ICASSP)}, 2020, pp. 486--490.

\end{thebibliography}

\end{document}